\documentclass[
  ,draft            
  ,numberedheadings 
  ]
  {aipproc}

\layoutstyle{8x11double}

\def\LCDM{$\Lambda$CDM~}
\def\LWDM{$\Lambda$WDM~}
\def\h2{H$_2$}
\def\ltsima{$\; \buildrel < \over \sim \;$}
\def\simlt{\lower.5ex\hbox{\ltsima}}
\def\gtsima{$\; \buildrel > \over \sim \;$}
\def\simgt{\lower.5ex\hbox{\gtsima}}

\def\cmm3{cm^{-3}}

\def\pp{$^2$\xspace}

\def\pop3{Pop~III\xspace}

\def\p3{``small-halo''\xspace}
\def\pp3{``Small-halo''\xspace}

\def\sig8{$\sigma_8$\xspace}

\newcommand{\hinv}{{h^{-1}}}
\newcommand{\himpc}{\hinv{\rm\,Mpc}}

\newcommand{\Msun}{\,\rm M_{\odot}}

\newcommand{\flw}{\ifmmode{F_{\rm{LW}}}\else{$F_{\rm{LW}}$}\fi}

\begin{document}

\title{Population III Star Formation and IMF}


\author{Michael L. Norman}{
  address={Center for Astrophysics and Space Sciences, University of California, San Diego, La Jolla, California USA; mlnorman@ucsd.edu}
}

\begin{abstract}
We review recent 3D cosmological hydrodynamic simulations of primordial star formation from cosmological initial conditions (Pop III.1) and from initial conditions that have been altered by radiative feedback from stellar sources (Pop III.2). We concentrate on simulations that resolve the formation of the gravitationally unstable cloud cores in mini-halos over the mass range $10^5 < M/\Msun < 10^7 $ and follow their evolution to densities of at least $10^{10} \cmm3$ and length scales of $<10^{-2}$ pc such that accretion rates can be estimated. The advent of ensembles of such simulations exploring a variety of conditions permits us to assess the robustness of the standard model for Pop III.1 star formation and investigate scatter in their formation redshifts and accretion rates, thereby providing much needed information about the Pop III IMF. The simulations confirm the prediction that Pop III.1 stars were massive ($\sim 100 \Msun$), and form in isolation in primordial mini-halos. Simulations of Pop III.2 star forming in relic HII regions suggest somewhat lower masses ($\sim 30 \Msun$) which may help explain the chemical abundances of extremely metal poor stars. We note that no 3D simulation at present has achieved stellar density let alone followed the entire accretion history of the star in any scenario, and thus the IMF of Pop III stars remains poorly determined theoretically. 
\end{abstract}

\maketitle


\section{The story up to 2003}
Dark matter mini-halos in the mass range $10^5-10^6$ solar mass virializing at high redshift are believed to collect enough primordial gas within them to host the formation of the first generation of stars (Population III). Mediated by molecular hydrogen cooling, gas in the centers of these halos condenses and becomes unstable to gravitational fragmentation with a typical mass scale of a few hundred solar masses. Pioneering high resolution 3D hydrodynamic cosmological simulations by Abel et al. (2002) and Bromm et al. (2002), and more recently by Yoshida et al. (2006) and O'Shea \& Norman (2007a) have shown that gravitational fragmentation produces only one massive fragment per halo. If this gas does not fragment further as it approaches stellar density, Pop III stars with this typical mass would be produced. The fates and feedback effects of such stars has many interesting consequences for the early structure formation as discussed elsewhere in this volume (see review by Ciardi). 

While this picture is now firmly established, based as it is on rather simple physics elucidated by the simulations, it is important to realize the remaining uncertainties. First, as of 2003 when the first reviews of primordial star formation were being written (Barkana \& Loeb 2001, Bromm \& Larson 2004, Glover 2005), a rather small number of such simulations had been done, raising the question of how ubiquitous this mode of star formation was in the early universe. Second, despite the large range of scales covered by the simulations, they were necessarily terminated well before stellar density was reached due to missing physics. The possibility of sub-fragmentation could not be ruled out, with consequent uncertainty on the mass scale of the first stars. Third, as emphasized by Barkana \& Loeb (2004), for technical reasons the simulations were carried out in quite small cosmological volumes (< 1 Mpc comoving), the result being that the redshift of formation of the first stars found ($z\sim 20$) was underestimated. Larger volumes would contain rarer peaks in the density field, and these would presumably form Pop III stars earlier (White \& Springel 2000). Could the higher densities and temperatures of this earlier epoch alter the mass scale of the truly first stars? And finally, there was a host of complicated feedback effects--radiative, kinetic, and chemical--which were not included in these first simulations which could ``mess up" the simple picture by altering the initial conditions and the cooling properties of the collapsing cloud cores.

The most concerning of these feedback effects was the build-up of an FUV background by emissions from the first stars that could photo-dissociate the hydrogen molecules which mediate Pop III star formation in the first place (Haiman, Rees \& Loeb 1997, Ciardi, Ferrara \& Abel 2000, Haiman, Abel \& Rees 2001). The first simulations examining this ``negative feedback effect" (Machacek, Bryan \& Abel 2001, hereafter MBA; and Yoshida et al 2003, hereafter Y03) concluded that Pop III star formation would not be suppressed, but merely delayed. The concept of a critical halo mass capable of forming a Pop III star was introduced by MBA. They found that the critical halo mass is an increasing function of the FUV background mean intensity in the Lyman-Werner bands (11.2-13.6 eV) suggesting that Pop III star formation would become self-limiting. Y03 found that gas cooling becomes inefficient above $J_{21}=0.01$ due to the photo-destruction of \h2. Here $J_{21}$ is the mean intensity of the FUV background in units of $10^{-21}$ ergs/cm$^2$/sec/ster/Hz. Using this information Y03 constructed a semi-analytic model of the cosmic history of Pop III star formation. They found a rapidly rising population of Pop III stars over the redshift interval 35 > z > 20 ``the rise of Pop III", which begins to become self-regulated at $z \sim 25$ when the FUV background attains $J_{21}=0.01$. They did not consider the lower redshift evolution as this was out of the range of their simulations, but rather speculated that Pop II star formation dominated the cosmic star formation history below $z \sim 20$. 

It is important to point out that neither the MBA simulations nor the Y03 simulations were evolved below redshifts of $z \sim 20$ or had the spatial resolution to follow the chemo-thermal-hydrodynamics of cloud collapse on scales below $\sim 0.1$ pc, making ``the fall of Pop III" by radiative feedbacks alone quite uncertain and understudied. Indeed, Y03 found the Pop III global SFR was still rising at the end of their simulations, implying that we don't even know when the Pop III epoch peaked. 

Ricotti, Gnedin \& Shull (2002) addressed this question using hydrodynamic cosmological simulations that included the feedback of both photo-dissociating and photo-ionizing radiation from primeval galaxies. Although their simulations were of even lower resolution than MBA and Y03 and parameterized star formation with ad hoc prescriptions, they found a cosmic star formation history quite similar to that predicted by Y03 despite small box size effects. They too could not simulate sufficiently large boxes to obtain converged results in the interesting redshift regime 20 > z > 6, but rather found that rare luminous objects strongly affected the average SFR within the simulation volume. It is fair to say that we have a much better idea when the first Pop III star formed in the universe than when the last one formed. Indeed, some models have been presented which predict Pop III stars forming as low as redshift 3 (Schneider et al. 2006). 

This review updates progress on direct numerical simulations of Pop III star formation by several groups since the reviews of Bromm and Larson (2004), Glover (2005), and Ciardi and Ferrara (2005). New high resolution simulations have been carried out by Yoshida et al. (2006), O'Shea \& Norman (2007a) and Gao et al. (2007) which test the robustness of the basic picture described above by pushing the collapse to higher central densities, although not yet stellar densities, and by considering ensembles of simulations. The effects of a FUV background on Pop III star formation has recently been revisited by O'Shea \& Norman (2007b) and by Wise and Abel (2007). It is found that \h2 cooling remains important in halos with masses approaching $10^8 \Msun$ and FUV backgrounds as strong as $J_{21}=1$. This suggests that the Pop III epoch may be more extended than previously thought and therefore occurring in rather different environments than originally simulated.  

Pop III star formation has been studied in new and different environments as well. Pop III star formation in warm dark matter models has been studied by O'Shea \& Norman (2006) and by Gao \& Theuns (2007). Finally, Pop III star formation in gas that has been pre-processed by earlier generation of Pop III stars has been simulated by O'Shea et al. (2005), Mesinger, Bryan \& Haiman (2006), and by Yoshida et al. (2007a,b). The surprising result is that despite the diversity of the environments studied, it is found that Pop III.2 stars form in basically the same way as Pop III.1 stars and that they are massive, although they appear to be somewhat less massive than Pop III.1 based on lower cloud core temperatures and accretion rates. However, the new simulations show that environment and formation redshift/history can have a substantial effect on protostellar accretions rates, suggesting some spread in the primordial IMF. 

No 3D simulation has yet been done that follows the entire accretion history of the star, and therefore final masses are still uncertain. However, approximate integrations have been done assuming spherical symmetry which suggest that the entire protostellar envelope can be accreted (Omukai \& Palla 2003, Yoshida et al. 2006, 2007a,b). Verifying this result with fully 3D simulations represents a grand challenge for the field. 
 
\section{Formation of Pop III.1 in $\Lambda$CDM: New Results}
Hereafter we will refer to Pop III stars forming from undisturbed cosmological initial conditions Pop III.1, in keeping with the taxonomy introduced by McKee (these proceedings). Our understanding of the formation of Pop III.1 stems primarily from high resolution 3D hydrodynamic cosmological simulations supplemented with primordial gas chemistry including the all-important reaction kinetics of molecular hydrogen. The simulations of Abel, Bryan \& Norman (2000, 2002) and Bromm, Coppi \& Larson (2002) showed that sufficient \h2 is formed in the cores of mini-halos of mass $10^5-10^6 \Msun$ at $z \sim 20$ so that ro-vibrational line cooling of \h2 will precipitate Jeans instability. Typical densities and temperatures in this gas prior to collapse are $n \sim 10^4 \cmm3$ and T $\sim$ 200 K, implying a Jeans mass of $\sim 500 \Msun$. 

\begin{figure}
  \includegraphics[height=.3\textheight]{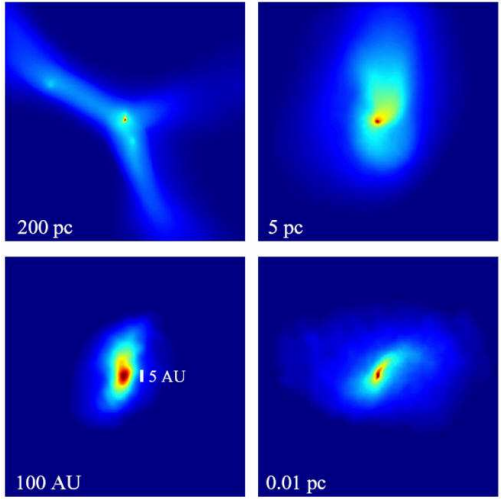}
  \caption{Formation of a primordial protostar using an ultra-high resolution cosmological SPH simulation. Projection of gas density on various scales. The central 1 pc region at top right contains $\sim 300 \Msun$ of gravitationally unstable primordial gas cooling by molecular hydrogen line radiation. The dark patch at lower left is a 0.01 $\Msun$ fully molecular hydrogen, optically thick cloud core cooling by collisionally induced emission with central density $3 \times 10^{15} \cmm3$. Accretion of the massive envelope onto this seed will form a Pop III star at $z \sim 19$. From Yoshida et. al. (2006).}
  \label{Yoshida06}
\end{figure}

The collapse of this gas initially proceeds quasi-statically, and then dynamically when the density threshold for 3-body \h2 formation is reached at $n \sim 10^8 \cmm3$. The collapse proceeds from the ``inside out", analogous to protostellar cloud collapse models that have been studied for decades in connection with current day star formation as reviewed by Larson (2003). The infalling envelope assumes a density distribution that is well approximated by a powerlaw: $n(r) \sim r^{-2.2}$ although the temperature distribution is distinctly not isothermal. Both groups found that the collapsing core did not fragment into smaller units, implying a massive star would eventually be formed. Both the Abel and Bromm simulations were forced to stop well before stellar density was reached due to limits of physics and resolution. The simulation by Abel, Bryan and Norman (2002) achieved a central density of $n \sim 10^{12} \cmm3$ on scales $\sim 100$ AU where a fully molecular core of $\sim 1 \Msun$ was found accreting its envelope at a rate of $\sim 10^{-2} \Msun$/yr. At this rate >100 $\Msun$ of material could accrete within a Kelvin time, suggesting that Pop III.1 stars were very massive. 

New simulations have confirmed and extended this basic picture. Table 1 summarizes the simulation details for the adaptive mesh refinement (AMR) simulation of O'Shea \& Norman (2007a; hereafter ON07a) and the smoothed particle hydrodynamics (SPH) simulation of Yoshida et al. (2006; hereafter Y06). Both groups assumed WMAP 1 cosmological parameters, had similar mass and spatial resolution, and ran their simulations until a central density of $10^{15}$ particles/cc was reached. The SPH simulation included more physics and is thus more reliable in the range of densities where radiation transport effects, collisionally-induced-emission (Ripamonti \& Abel 2004), and equation of state effects become important ($n > 10^{10} \cmm3$). Y06 ran a single simulation in a box 0.3 Mpc on a side, while ON07a ran a dozen simulations in three different box sizes: 0.3, 0.45 and 0.6 $\himpc$. We will therefore use the ON07a simulations to discuss the effect of formation redshift and environment, the Y06 simulation to discuss high density evolution, and compare the two of them over the range of densities where they can be compared.

\begin{table}
\begin{tabular}{|l|l|l|}
\hline
   & \emph{AMR} & \emph{SPH} \\
\hline
\hline
Code & Enzo & GADGET \\
\hline
Cosmological model & WMAP1 & WMAP1 \\
\hline
Box size ($\himpc$) & 0.3, 0.45, 0.6 & 0.21 \\
\hline
Simulations & 12 & 1 \\
\hline 
$m_{dm} (\Msun)$ & 2.6, 8.8, 21 & 0.1 \\
\hline
$m_{gas} (\Msun)$ & $0.4 \rightarrow 5 \times 10^{-4}$ & 0.015 \\
\hline
$n_{max} (\cmm3)$ & $10^{15}$ & $10^{15}$ \\
\hline
Physics & H, He, \h2 chemistry & H, He, \h2, HD chemistry\\
        & optically thin cooling  & \h2, HD, and CIE cooling\\
        & (\h2 only)             & radiative transfer \\
\hline
Minimum length & $3 \times 10^{-6}$ & $3 \times 10^{-6}$ \\
scale (pc) & & \\
\hline
Minimum mass & $10^{-3}$ & $10^{-2}$ \\
scale ($\Msun$) & &\\
\hline
\hline

\end{tabular}
\caption{Simulation parameters}
\end{table}

\subsection{Comparison of AMR and SPH results}
Fig. 1 shows the radial density and temperature profiles when the central density has reached $10^{15} \cmm3$. Although the ON07a simulation is not valid at these high densities, the density profiles agree extremely well over a large range of radii and enclosed masses considering the very different numerical techniques employed. The density profile is well approximated as power law $n \sim r^{-2.2}$ over many decades in radius. The temperature distributions also agree extremely well for n < $10^{10} \cmm3$ where opacity effects become important. The temperature distribution reflects four regions. Proceeding from larger to smaller radii we have first the region of cosmological infall, where the temperature increases to its maximum value ($\sim 1000$K) at the virialization shock located at $r \sim 100$ pc. The second is a cooling zone where the gas temperature drops to the minimum allowed by \h2 cooling ($\sim 200$ K) at a radius of a few pc. The third sees the gas temperature rise again to $\sim 1000$K due to compressional heating in the quasi-statically collapsing cloud core at $r \sim .01$ pc. The fourth and innermost ($r < 10^{-3}$ pc) is a cooling zone accompanying the formation of the fully molecular core due to the 3-body \h2 reaction. This innermost cooling zone is absent in the Y06 simulation because the cooling radiation is trapped at densities $n > 10^{10} \cmm3$, whereas it is allowed to escape unphysically in the ON07a simulation. When opacity effects are included, the central temperature rises to $\sim 2000$ K. The difference in thermal structure seems to have little effect on the density profile, however.

\begin{figure}
  \includegraphics[width=.7\textwidth]{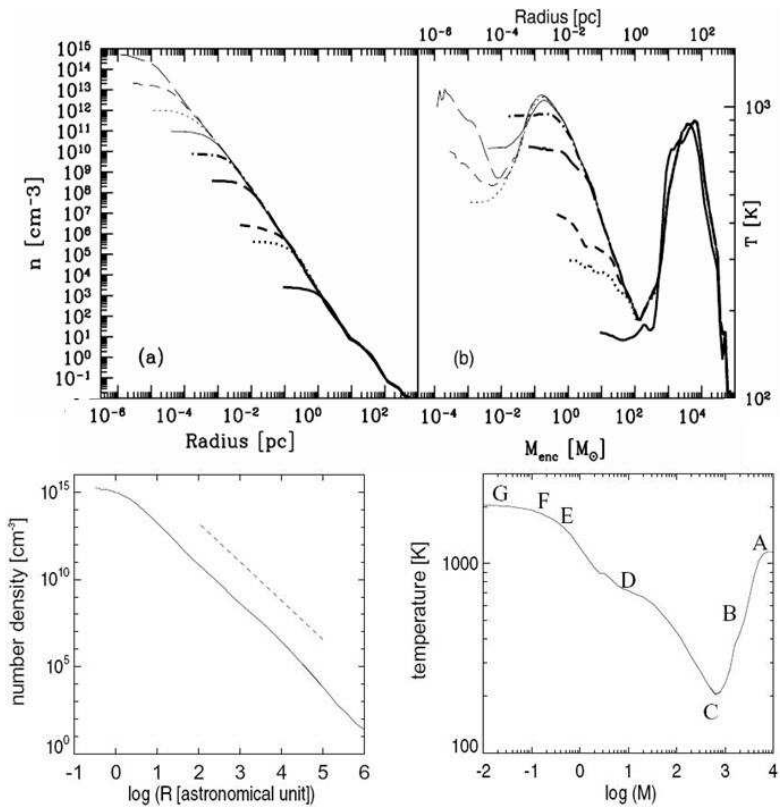}
  \caption{Radial density and temperature structure of Pop III star forming cloud, as found from AMR (O'Shea \& Norman 2007a) and SPH (Yoshida et al. 2006) numerical simulations. Note the excellent agreement for $n < 10^{10} \cmm3 (r > 10^{-3}$ pc) where the gas is optically thin to \h2 cooling radiation.}
  \label{AMRvsSPH}
\end{figure}

\begin{figure}
  \includegraphics[width=.7\textwidth]{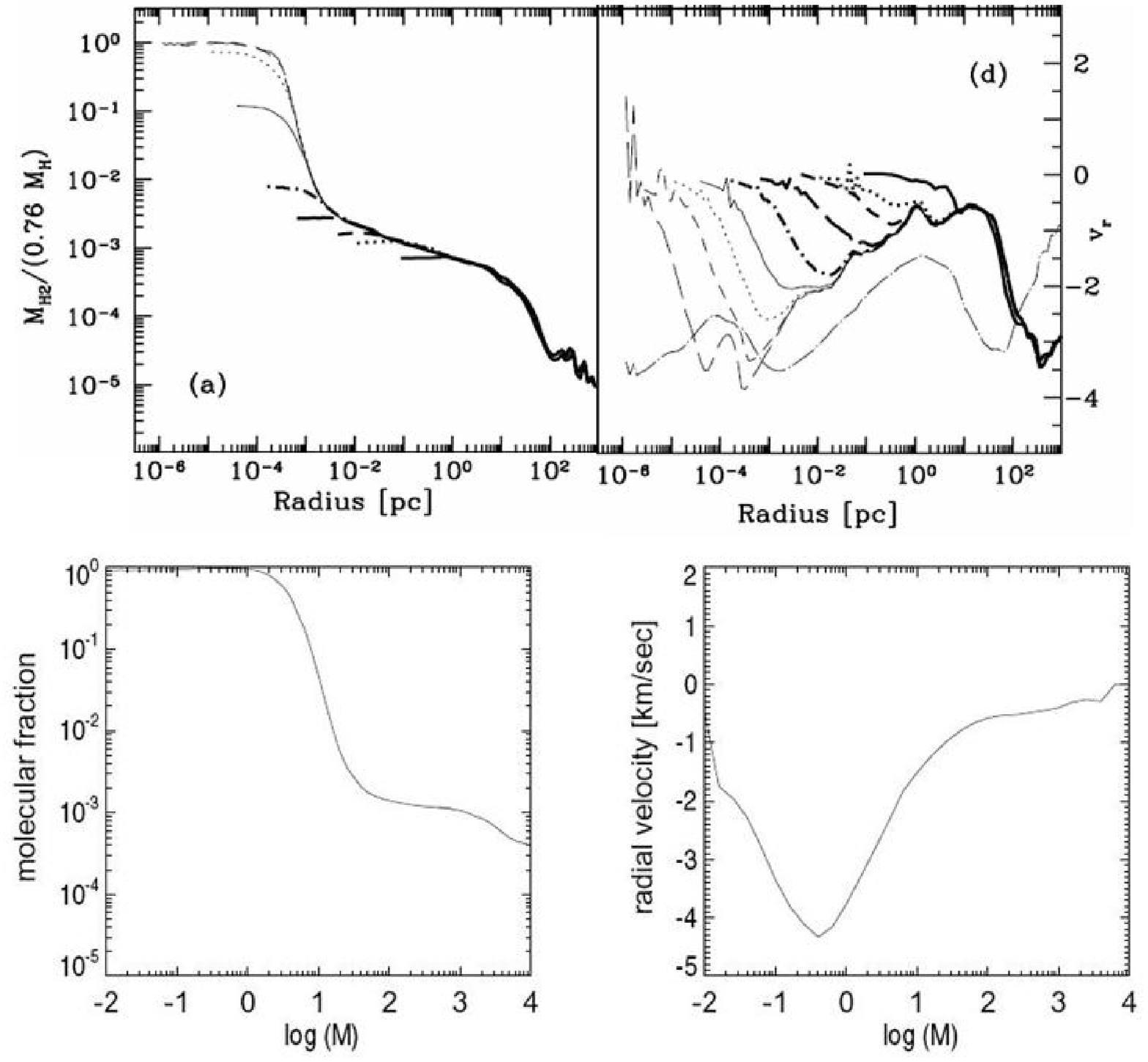}
  \caption{Radial profiles of $H_2$ fraction and radial velocity from the AMR and SPH simulations referred to in Fig. 2.}
  \label{AMRvsSP\h2}
\end{figure}

Fig. 3 shows a comparison of profiles for molecular hydrogen fraction and radial velocity, analogous to Fig. 2. Three body \h2 formation creates a fully molecular core of size $10^{-4}$ pc and mass $\sim 1 \Msun$. The massive envelope accretes onto this seed, driving its thermal and chemical evolution. A maximum infall velocity of  $-4$ km/s is achieved in both simulations at a radius of $\sim 10^{-4}$ pc.

The collapse of the cloud core is relatively unimpeded by angular momentum effects. Fig. 4 shows the specific angular momentum profiles from both simulations. Again, the agreement is excellent both qualitatively and quantitatively. The star forming cloud has a sharply increasing specific angular momentum distribution as a function of enclosed mass, which is well-approximated by $L(M)\propto M$. This implies the first stars form from very low angular momentum material. A plot of the ratio of the circular velocity to the local Keplerian velocity on mass shells (dashed line of Fig. 4b) indicates that the infalling envelope is never rotationally supported. 

How is this angular momentum profile established? The multiple lines plotted in Fig. 4a indicate different output times corresponding to the density profiles shown in Fig. 2. The change with time of the angular momentum profile implies redistribution of angular momentum within the innermost 1000 $\Msun$ of the cloud. This is seen in both AMR and SPH simulations, and is understood to be the result of angular momentum segregation, wherein lower angular momentum fluid elements in the turbulent cloud preferentially sink to the center (O'Shea 2005, Y06).

\begin{figure}
  \includegraphics[width=.4\textwidth]{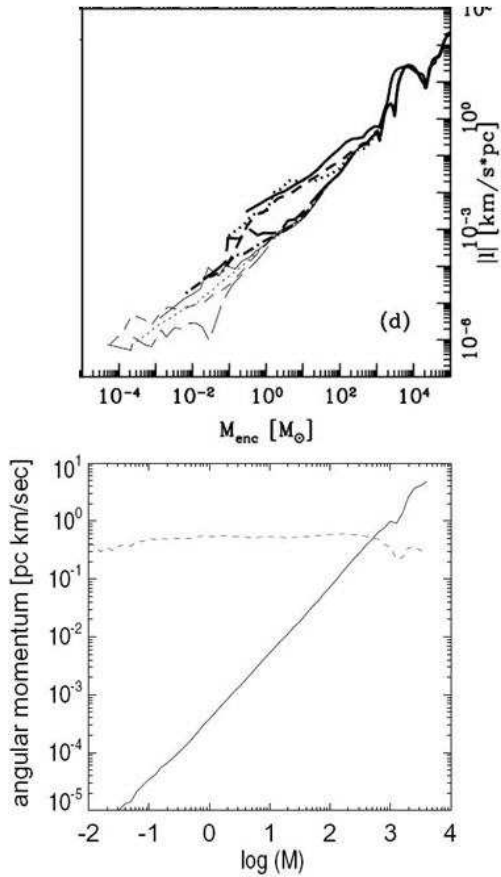}
  \caption{Specific angular momentum profiles in Pop III star forming cloud. Top: time evolution in the AMR simulation. Bottom: final time in the SPH simulation}
  \label{angmom}
\end{figure}

\subsection{Evolution to higher densities: fragmentation?}
At densities above $10^8 \cmm3$ 3-body formation of \h2 becomes efficient at converting the mostly atomic gas into molecular hydrogen. At densities above $10^{10} \cmm3$ it is necessary to include the trapping of the \h2 line radiation, collisionally induced emission (CIE), and chemical heating of the gas as the core becomes fully molecular. These effects have been included in the simulation of Y06, and therefore we refer to their results in the following. 

The effect of radiation trapping and chemical heating is to raise the temperature of the core to $\sim 2000$ K as n reaches $10^{15} \cmm3$. Prior to reaching these densities and temperatures the core passes through a chemo-thermally (CT) unstable regime which Silk (1983) calculated would lead to fragmentation. Y06 evaluated the growth time of the CT instability using the conditions in the center of their cloud, and found that it briefly becomes comparable to the dynamical time at $n \sim 10^{10} \cmm3$ (Fig. 5), but that there is insufficient material satisfying these conditions to form multiple fragments. 

CIE becomes the dominant coolant above $n \sim 10^{14} \cmm3$, however the dynamical timescale in the core remains short compared to the timescale for isobaric modes of the thermal instability, and therefore the core does not fragment into multiple objects. As the central density and temperature increases, the Jeans mass continues to drop, suggesting low mass stars could be produced if a fragmentation mechanism existed that operated quickly enough. However the simulations and the linear analysis reveals that multiple fragments are not produced, but rather a single fragment which is the accretion center for the rest of the cloud. Abel refers to this process as ``Russian doll fragmentation." This result needs to be kept in mind when interpreting the claims that very low mass stars can results from dust-induced fragmentation (Ciardi, these proceedings). 

\begin{figure}
  \includegraphics[width=.4\textwidth]{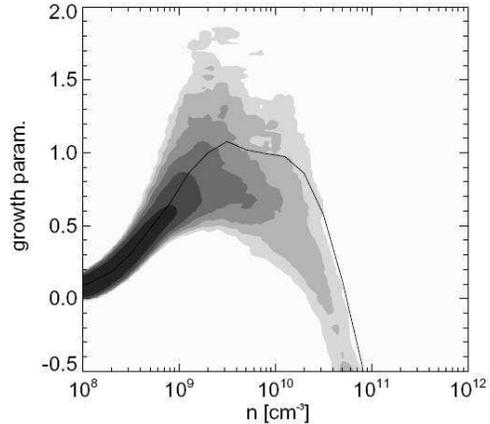}
  \caption{Evolution of chemo-thermal instability in the collapsing Pop III cloud core. A growth parameter >2 is required for the core to fragment into multiple pieces. The core does not fragment, but produces a single protostellar seed which accretes the massive envelope. See Yoshida et al. (2006) for more details.}
  \label{growth}
\end{figure}

\subsection{Evolution of the protostar}
3D simulations have not yet been able to evolve the cloud core to stellar densities, although valiant attempts are being made as reported by Yoshida and Turk at this meeting. Additional physical processes must be included, such as collisional dissociation of \h2 and the trapping of continuum radiation due to the CIE opacity. The further evolution of the core depends critically on the role of angular momentum in halting core collapse and building a centrifugally supported disk through which most of the envelope mass would presumably be accreted (Tan \& McKee 2004). In the absence of rotation, we have the 1D spherically symmetric models of Omukai and Nishi (1999) and Omukai \& Palla (2003), which show that the entire envelope can be accreted provided the late time accretion rate does not exceed the Eddington rate of $\dot{m} = 4 \times 10^{-3} \Msun$/yr. A plot of the instantaneous accretion rate at the end of the Y06 simulation is shown in Fig. 6-top. The accretion rate drops below the Eddington-limited rate after 10 $\Msun$ has been accreted, just as the protostar begins its Kelvin-Helmholtz contraction to the ZAMS (Fig. 6-bottom). Thus most of the protostar's pre-main sequence evolution is sub-Eddington, implying that rotation, not radiation, sets its ZAMS mass. The three lines in Fig. 6b show protostellar evolution tracks for different assumptions about how much centrifugal forces reduces the accretion rate onto the protostar. For even quite significant reductions in the accretion rate by rotation, the Pop III star is predicted to enter the main sequence with >60 $\Msun$. 

\begin{figure}
  \includegraphics[width=.4\textwidth]{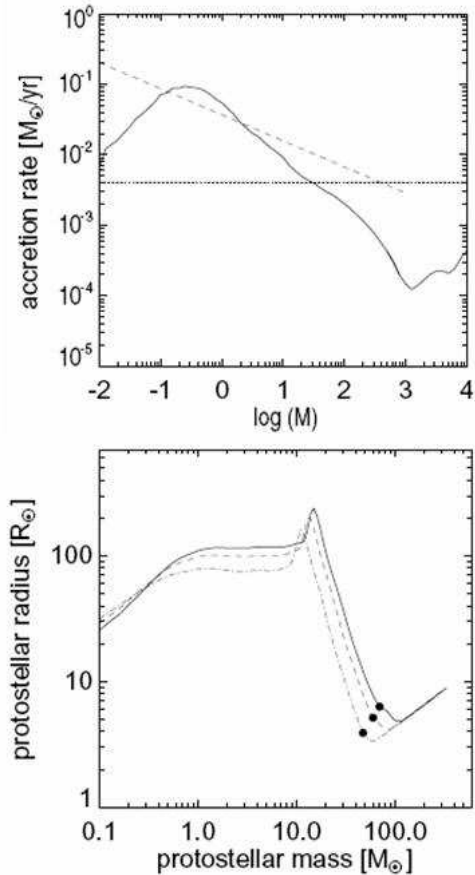}
  \caption{Top: Instantaneous accretion rate onto the protostar depicted in Fig. 1. Bottom: Inferred evolution of the protostar based on a spherically-symmetric evolution model for 1, 2/3, and 1/3 the accretion rate shown above. Dots indicate cessation of contraction when CNO cycle hydrogen burning begins. From Yoshida et al. (2006).}
  \label{protostar}
\end{figure} 

\subsection{Dependence on formation redshift and environment}
ON07a and Gao et al. (2007) have studied the dependence of the central protostellar cloud's properties on formation redshift and cosmological environment. As pointed out by White and Springel (2000), dark matter mini-halos in the mass range capable of forming Pop III.1 would be expected to form in ``protocluster" environments at redshifts considerably higher than 20. A pure dark matter N-body study of such an environment by Gao et al. (2005) found that formation redshifts as high as 50 are likely (assuming $\sigma_8=0.9$). In this case, the higher densities and temperatures in the collapsing cloud would shorten formation timescales and increase accretion rates. ON07a simulated 12 cases by varying box size and random seed assuming a WMAP1 cosmological parameters. Their sample of halos formed stars over the redshift interval 33 > z > 19. Gao et al. (2007) simulated 8 cases using a multimass technique whereby a select region from a much larger cosmological volume is re-simulated at much higher mass and spatial resolution. They considered WMAP1 and WMAP3 cosmological parameters (differing principally in the amount of small-scale power), and also different baryon fractions. They were thus able to study halos forming stars over the larger redshift interval 50 > z >10. Both simulations terminated their evolutions when central densities of $10^{10} \cmm3$ were reached and ignored the effects of a photo-dissociating UV background and opacity effects in the collapsing cloud core. 

\begin{figure}
  \includegraphics[width=.4\textwidth]{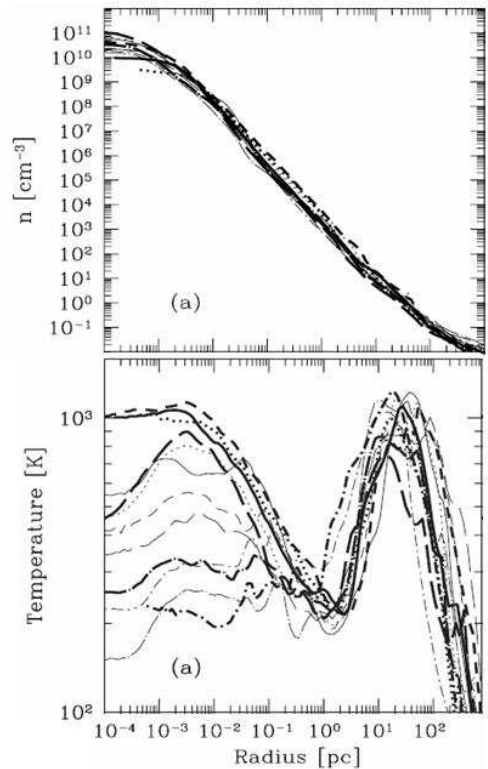}
  \caption{Density and temperature profiles for an ensemble of 12 AMR simulations when central densities have reached $10^{10} cm^{-3}$. Density profile is insensitive for formation redshift and environment, while temperature profile shows considerable variation in the central 1 pc. From O'Shea \& Norman (2007a). }
  \label{dentem12}
\end{figure} 
 
Fig. 7 shows the density and temperature distributions at the end of the 12 simulations in ON07a. The density profiles are very similar to one another despite differences in formation redshift and environment, while the temperature profiles show a considerable amount of variation in the core region. 
This is reflected in the instantaneous accretion rate plots (Fig. 8).  While all 12 simulations show the generic feature of an early rapid rise to a maximum followed by a power-law decline, there is a two order of magnitude scatter in peak accretion rates. ON07a found a correlation between redshift of collapse (which correlates with box size) and peak accretion rate, with higher redshift minihalos exhibiting lower accretion rates. This was understood to be the result of the early rapid formation of \h2 in the denser, hotter high redshift mini-halos which results in lower final core temperatures and hence accretion rates, since $\dot{m} \sim c_s^3/G$. 

Gao et al. (2007) also found considerable scatter in accretion rates, but no redshift correlation. Rather, they found their cores had a variety of shapes and amount of centrifugal support, and that their centrifugally-supported disk-like cores had the lowest accretion rates. Clearly, both of these effects could be operating and bear further study. Given the importance of the late-time accretion rates on setting the stellar mass, a careful study of angular momentum evolution in the cloud cores is needed.

\begin{figure}
  \includegraphics[width=.4\textwidth]{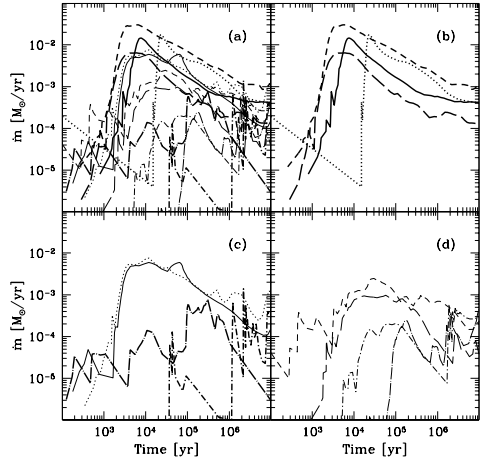}
  \caption{Implied accretion rate versus time from an ensemble of 12 AMR simulations of Pop III star formation, as inferred from the instantaneous accretion rate at the end of each simulation (when central densities reached $10^{10} \cmm3$). From O'Shea \& Norman (2007a). }
  \label{accretion12}
\end{figure}

\section{Formation of Pop III.1 in $\Lambda$WDM}
O'Shea \& Norman (2006; hereafter ON06) and Gao \& Theuns (2007; hereafter GT07) have considered the formation of Pop III star in a universe whose dominant mass component is warm dark matter (WDM). WDM has been advanced as a solution to the many problems of CDM on sub-Mpc scales (e.g., Bode, Ostriker \& Turok 2001). Operationally, WDM simulations are similar to CDM simulations except that the CDM power spectrum is assumed to by exponentially suppressed below some scale set by the free streaming scale of the WDM particle, which depends upon its mass for thermal relics. Fig. 9 shows the suppression mass scale for gravitino WDM, given by 

\begin{equation}
M_{supp} = 10^{10} \left( \frac{\Omega_{WDM}}{0.3} \right)^{1.45}  
\left( \frac{h}{0.65} \right)^{3.9} \left( \frac{ 1~keV }{ m_{WDM} } \right)^{3.45}   h^{-1} M_\odot
\end{equation}

As can be seen, the mini-halos which form Pop III.1 stars in \LCDM become suppressed for $m_{WDM}$<20 kev, raising the interesting question of what is the nature of the first cosmological objects in \LWDM?

\begin{figure}
  \includegraphics[width=.4\textwidth]{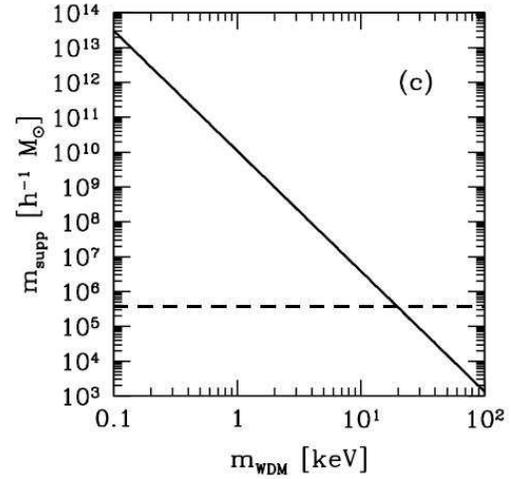}
  \caption{Effects of gravitino warm dark matter on the mass scale where perturbations are suppressed by free streaming. Cosmological mini-halos hosting the formation of Pop III stars become suppressed for $m_{WDM}$ < 20 keV. From O'Shea \& Norman (2006).  }
  \label{msuppression}
\end{figure}

ON06 carried out 8 high resolution AMR simulations with resolutions similar to Table 1 however varying the WDM particle mass over the range $12.5 \leq m_{WDM} \leq 40$ keV. They followed the evolution until the baryonic core had collapsed to baryon densities of $n=10^{10} \cmm3$. Fig. 10 shows the collapse redshift as a function of $m_{WDM}$. ON06 found that the $m_{WDM}$=40 keV evolution is virtually indistinguishable from the \LCDM case computed with the same random seed, which collapsed at z=17.8. The effects of decreasing the WDM particle mass is to modify the dark matter distribution at virialization from spheroidal halos to elongated filaments (Bode, Ostriker \& Turok 2001) and to reduce the core collapse redshift. The $m_{WDM}$=12.5 keV case produces a collapsing primordial cloud core at z=12 in the center of a cosmological filament (Fig. 11) which is considerably more massive ($\sim 10^7 \Msun$) than its CDM counterpart. Despite these differences, it is found that the high density ($n>10^5 \cmm3$) evolution of the central $\sim 500 \Msun$ of the collapsing cloud core is very similar to the CDM case (Fig. 12), implying that similarly massive stars would be produced. 

\begin{figure}
  \includegraphics[width=.4\textwidth]{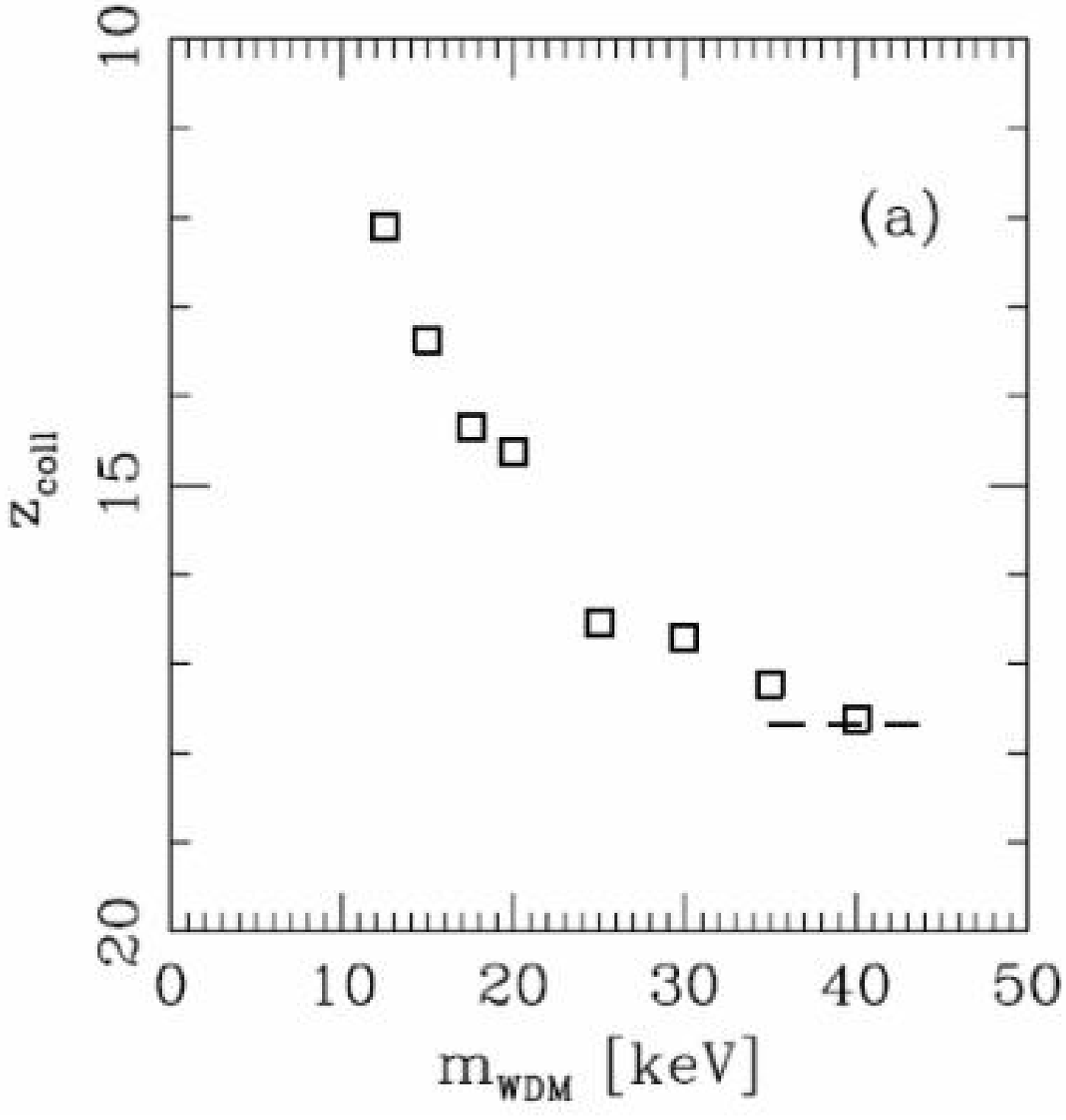}=0.01
  \caption{Collapse redhsift versus WDM particle mass for gravitationally unstable primordial cloud core. From O'Shea \& Norman (2006).}
  \label{zcollapse}
\end{figure}

A different conclusion was reached by GT07, who simulated the more extreme case $m_{WDM}$=3 keV. In this case, the suppression mass scale is $\sim 10^8 \Msun$. Objects at this mass scale virialize into long ($\sim 100$ kpc comoving) slender filaments with no dark matter substructure around which the baryons coalesce. Their simulation was terminated when the filament achieved a central density of $n \sim 10^5 \cmm3$ and temperature $T \sim 200$ K, but before it fragments. Presumably gravitational instability would fragment the filament into lumps of order the central Jeans mass which is a few hundred solar masses for the values above. Yet they state otherwise. They claim the filament will continue to collapse to higher densities in analogy to \LCDM cloud cores, and thereby fragment into lower mass protostellar seeds. They suggest the seeds will then accrete from the cylindrical gaseous envelope while they merge along the filament axis, producing a broad distribution of Pop III stellar masses. This scenario needs to be investigated with high resolution numerical simulations because as we have seen, whether the cloud fragments or not depends not only on the local Jeans mass, but on a competition of timescales which can only be determined from a self-consistent multidimensional calculation. 

\begin{figure}
  \includegraphics[width=.4\textwidth]{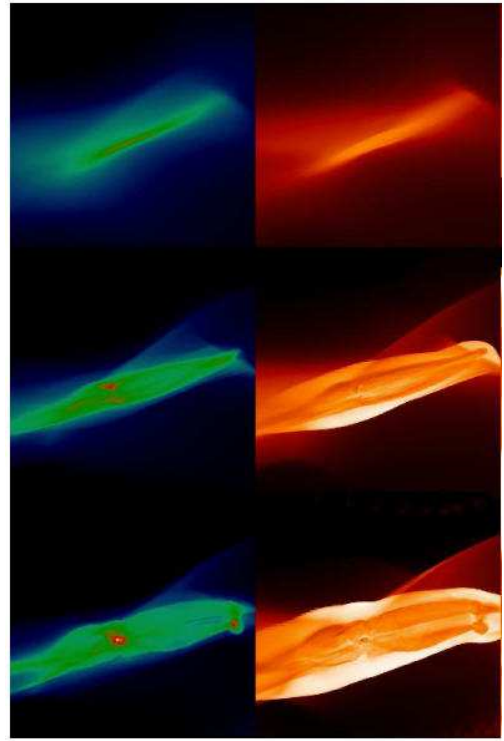}
  \caption{Formation of Pop III star in a cosmological filament in a \LWDM universe. From O'Shea \& Norman (2006)}
  \label{filament}
\end{figure} 

\begin{figure}
  \includegraphics[width=.4\textwidth]{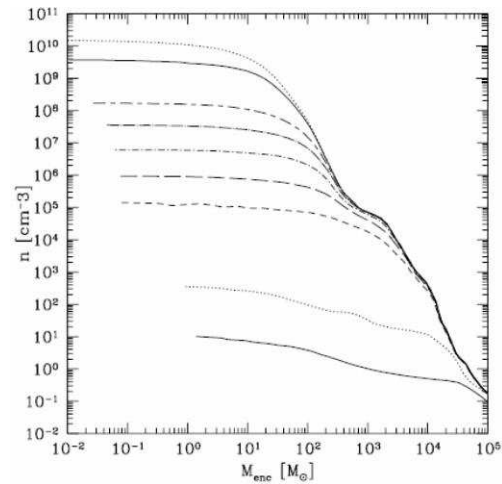}
  \caption{Evolution of density profile in the collapsing cloud core in the $m_{WDM}$=12.5 kev simulation shown in Fig. 11. For $n > 10^5 \cmm3$ the $\sim 500 \Msun$ core evolves very similarly to the \LCDM case despite the delay in achieving collapse conditions. From O'Shea \& Norman (2006). }
  \label{denevol}
\end{figure} 

\section{Formation of Pop III.2: Variations on a theme}
In this section we return to simulations of the formation of Pop III stars in a \LCDM universe which examine the influence of radiation backgrounds on their formation history. We discuss two cases in particular, the formation in the presence of a concurrent photo-dissociating Lyman-Werner background, and formation from gas that has previously been ionized by stellar sources (relic HII regions). 

\subsection{Effect of a Lyman-Werner UV background}
Machacek, Bryan \& Abel (2001) showed using moderate resolution AMR survey simulations that a Lyman-Werner background radiation field delays but does not extinguish Pop III star formation by raising the minimum halo mass capable of cooling by \h2 line transitions. They carried out simulations for fixed LW backgrounds corresponding to $J_{21}=0, 10^{-2.1}, 10^{-1.1}$. The formula for a fit to their minimum halo mass is given by Ciardi in these proceedings, but generally rises from $M_{min}=10^5 \Msun$ at $J_{21}$=0 to $M_{min}=10^6 \Msun$ at $J_{21}=10^{-1.1}$. Wise and Abel (2005) extrapolated the fit to LW backgrounds as high as $J_{21}$=1 and used a Press-Schechter-type analysis to estimate the evolution of the LW background for three values of the assumed mass of the the Pop III stars produced. They found that Pop III star formation becomes self limiting by z=25 and that the LW background reaches $J_{21}=0.1-1$ in the redshift range 20 > z > 10. Y03 carried out SPH survey simulations similar in resolution to those of MBA and found that \h2 cooling becomes inefficient in primordial mini-halos for $J_{21}$>0.01, making the extrapolation of the MBA minimum halo mass fitting formula to $J_{21}$=1 somewhat suspect. Neither the MBA nor the Y03 simulations had the mass or spatial resolution to simulate baryonic core collapse. Motivated by this, O'Shea \& Norman (2007b) carried out such simulations.

Fig. 13 shows the time evolution of central values for gas density, temperature, entropy, and \h2 fraction for four values of $J_{21}$. The figures show the time delay effect very clearly. The physics is very simple. For $J_{21}$>0.01 the photodissocation time is shorter than the 2-body formation time, consequently the \h2 fraction assumes its equilibrium value. The cooling time is then proportional to $J_{21}$, and reaches $10^8$ yr for $J_{21}$=1. The surprising result is that collapse eventually occurs even in the extreme case $J_{21}$=1 despite the greatly reduced \h2 fraction in the core. 

The explanation for this result is that the time delay allows the virial mass and temperature of the halo to increase through mergers to $M_{vir} \sim 10^7 \Msun$ and $T_{vir} \sim 2000$ K where the per molecule cooling rate in the core is greatly elevated. In the most extreme case, central temperatures reach 5000 K which accelerates production and cooling of \h2 to the point of baryonic collapse. Fig. 14-top shows the increase of the mass-weighted temperature in the central $100 \Msun$ of gas at the collapse redshift for increasing values of $J_{21}$. This translates into larger accretion rates onto the protostar at early times (Fig. 14-bottom) exceeding $10^{-2} \Msun$/yr for $J_{21}$>0.01. Taken at face value, this suggests that Pop III.2 stars forming in high LW backgrounds may be more massive than Pop III.1 stars, although we must bear in mind that it is the late-time accretion rate that governs the final mass of the star. This suggestion bears further study. 

\begin{figure}
  \includegraphics[width=.4\textwidth]{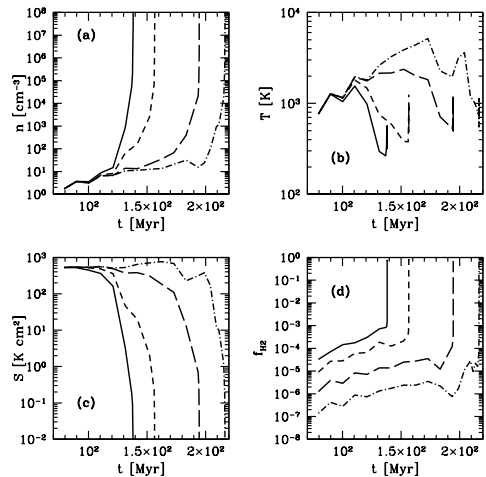}
  \caption{Evolution of central density, temperature, entropy, and \h2 fraction in a mini-halo for different Lyman-Werner UV background intensities (solid: $J_{21}=10^{-3}$, short dash: $J_{21}=10^{-2}$, long dash: $J_{21}=10^{-1}$, dot dash: $J_{21}=1$). In the strongest UVB, gravitational collapse is delayed by 100 Myr, but not prevented. From O'Shea \& Norman (2007b). }
  \label{LWcentral}
\end{figure}  

\begin{figure}
  \includegraphics[width=.4\textwidth]{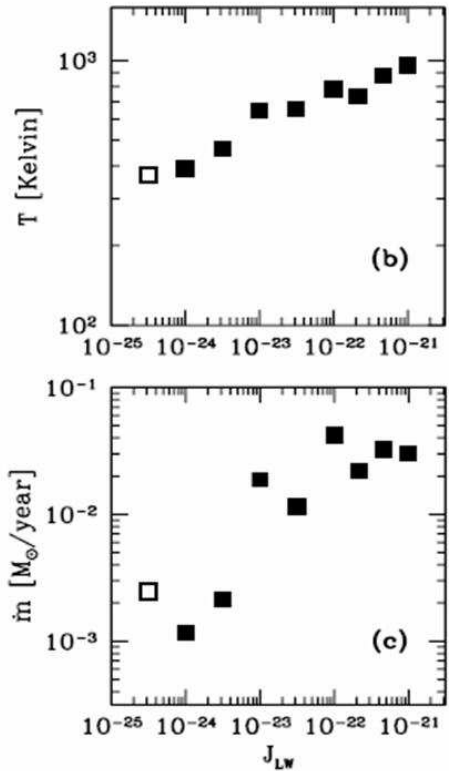}
  \caption{Conditions in the central 100 $\Msun$ cloud core as a function of LW mean intensity. Top: mean temperature, Bottom: accretion rate. Because a LW background delays baryonic collapse, virial temperatures, core temperatures and hence accretion rates are elevated in late-forming primordial mini-halos. From O'Shea \& Norman (2007b). }
  \label{LWtrends}
\end{figure}  

Wise \& Abel (2007) independently carried out a study similar to O'Shea \& Norman (2007b), but considered even more massive halos ($M_{vir} \leq 4 x 10^7 \Msun$) where \h2 cooling gives way the Lyman alpha cooling. They find a smooth transition in collapse redshift and halo mass between those that cool by \h2 lines and those that cool by atomic hydrogen lines (Fig. 15). They find that collisional ionization of H in the warmest halos produces enough electrons to catalyze the formation of \h2 even in the presence of strong LW backgrounds. The conclusion from both of these works is that \h2 cooling remains the dominant mediator of early star formation even for strong UVBs, and that by restricting ones attention to only those halos for which $T_{vir} >10^4$ K one misses a large fraction of star forming halos. 

\begin{figure}
  \includegraphics[width=.4\textwidth]{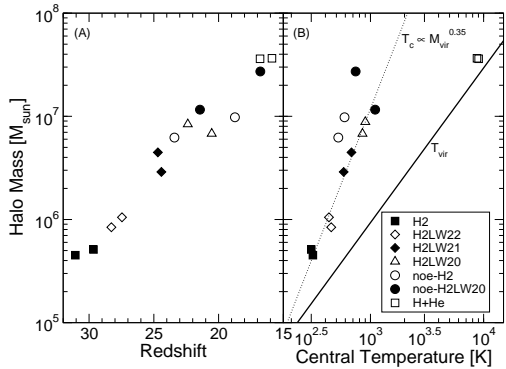}
  \caption{Smooth transition from \h2 coolers to Lyman alpha coolers, in terms of collapse redshift and central temperature. From Wise \& Abel (2007).}
  \label{WiseAbel}
\end{figure}

\subsection{Formation in a relic HII region}
Finally we consider the case of a mini-halo that has previously been ionized by a strong UV source, Pop III or otherwise. As reviewed by Ciardi (these proceedings), Pop III stars photo-evaporate most of the baryons from the halos that formed them, and create large HII regions of 1-2 kpc in diameter. After a few Myr the Pop III star dies, leaving behind a relic HII region. Gas cools and recombines out of equilibrium, and the large residual electron fraction quickly catalyzes the formation of \h2 via the H- reaction (O'Shea et al. 2005). The question is under what conditions does a neighbor halo form a primordial star, and in those that do, what are their properties?

The survival of neighbor halos depends on a number of factors including halo mass, evolutionary stage, proximity to the ionizing source, and its ionizing luminosity (Ahn \& Shapiro 2007, Whalen et al. 2007). The most important factor appears to be the central density of the neighboring halo at the time of the time it is illuminated by the ionizing source. Densities of $n \sim 10^4 \cmm3$, corresponding to the ``loitering phase" of the evolution of primordial cloud cores, are high enough to self-shield the cloud against the LW radiation from the first star. Cores of lower density are photo-dissociated, but crushed to higher density by the cometary I-front. The residual ionization in these cores rapidly produces \h2 after the first star dies. Only the lowest density neighbor halos have their baryons stripped, and are therefore unable to form stars. Overall, the effect on star formation appears to be neutral (Ahn \& Shapiro 2007). 

O'Shea et al. (2005) carried out a high resolution 3D AMR cosmological simulation of the evolution of a halo forming inside the relic HII region of a Pop III.1 star. Resolutions similar to that given in Table 1 were used to follow the collapse of the second halo, which began 23 Myr after the first star died. The collapse proceeded in a way very similar to that of a Pop III.1 star once central densities of $10^4 \cmm3$ were reached, indicating the universality of how Pop III stars are formed. However, the instantaneous accretion rate when central densities of $10^{10} \cmm3$ were reached was quite low ($\dot{m} \sim 10^{-4} \Msun$/yr) due to the fact that the second halo had a large amount of angular momentum and produced a centrifugally-supported disk in its center. 

\subsection{Effect of preionization and HD cooling}
So far we have considered only Pop III stars forming from gas that has remained neutral since recombination. Stars forming in chemically pristine gas that has previously been ionized start from initial conditions that differ thermally, chemically, structurally, and kinematically from Pop III.1. Photoionization will raise the entropy of the gas and smooth out the clumping due to photoheating and photo-evaporation (Oh \& Haiman 2003). In addition, the high electron fraction will catalyze rapid formation of \h2 molecules via the H- channel when the HII region begins to recombine after the ionizing source switches off (O'Shea et al. 2005). In the absence of a high LW background, \h2 fractions are elevated by 2 orders of magnitude at the cosmic mean density relative to the Pop III.1 case. Finally, these conditions are ideal for the formation of HD, which becomes an important coolant at moderate densities ($10 < n < 10^6 \cmm3$) and low temperatures (T<150K) due to chemical fractionation and HD's large permanent dipole moment (Glover, these proceedings). 

The effect of these processes on the evolution of Pop III star-forming halos has been examined by Mesinger, Bryan \& Haiman (2006) (see also Bryan et al., these proceedings). They carried out high resolution AMR simulations using Enzo of a region that forms a Pop III.1 star at z=25 in the absence of radiative feedback. Without HD cooling, they found that the negative and positive feedback effects roughly cancel, with negative feedback (density smoothing) slightly outweighing positive feedback (accelerated \h2 formation). Pop III.2 stars formed in such regions are found to have very similar properties as Pop III.1 stars as estimated by mass accretion rates. When HD cooling is included, however, the simulations predict a reduced mass accretion rate because of the lower tempersture and hence soundspeed in the accreting gas. Pop III.2 stars with typical masses of $30-40 \Msun$ are claimed. 

\subsection{Second generation primordial stars}
A similar conclusion was reached by Yoshida et al. (2007a,b), who recently have simulated the second primordial star to form in a given primordial mini-halo. They self-consistently simulate the formation of a Pop III.1 star as in Y06, and follow the expansion of the HII/HeIII region it produces. As in earlier studies (Whalen et al. 2004, Kitiyama et al. 2004), gas in the minihalo core is swept out to the virial radius by a strong photo-evaporative flow. After the Pop III.1 star dies, this gas recombines and cools, forming HD. They continue the calculation until this gas recollects into the dark matter mini-halo core and forms a Pop III.2 star. They find that it takes about 100 Myr for the second primordial star to form, and that as a consequence of HD cooling, the mass accretion rate is reduced, in agreement with the results of Mesinger, Bryan \& Haiman (2006). In Yoshida, Omukai \& Hernquist (2007) they calculate the accretion history of the Pop III.2 protostar using a 1D protostellar accretion code. The result is shown in Fig. 16. It enters the main sequence with a mass of $\sim 30 \Msun$, significantly less than the Pop III.1 star that pre-processed its birth cloud.

\begin{figure}
  \includegraphics[width=.4\textwidth]{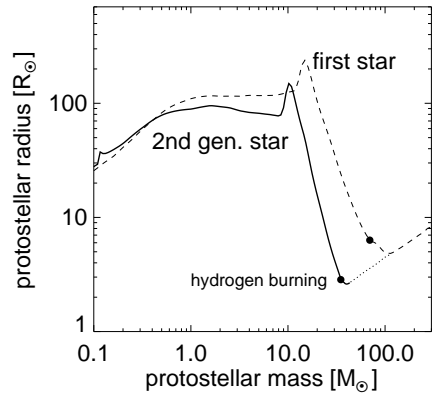}
  \caption{Evolution of the protostellar mass and radius for a Pop III.2 star forming in the halo of an earlier Pop III.1 star which photoionized its environment. From Yoshida, Omukai \& Hernquist (2007).}
  \label{2ndprotostar}
\end{figure}


\begin{theacknowledgments}
MLN wishes to thank the conference organizers for their financial support for attending the meeting in Santa Fe, and their patience while this manuscript
was being prepared. MLN also wishes to thank Tom Abel, Brian O'Shea, and Naoki Yoshida for permission to reproduce their figures. 
The research of MLN is supported by the National Science Foundation through
grant AST-0307690.

\end{theacknowledgments}




\end{document}